\documentclass[conference]{IEEEtran}
\IEEEoverridecommandlockouts
\usepackage{cite}
\usepackage{amsmath,amssymb,amsfonts}
\usepackage{algorithmic}
\usepackage{graphicx}
\usepackage{textcomp}
\usepackage{xcolor}
\usepackage{booktabs}       
\usepackage{algorithm}
\usepackage{wrapfig}

\usepackage{caption}
\def\BibTeX{{\rm B\kern-.05em{\sc i\kern-.025em b}\kern-.08em
    T\kern-.1667em\lower.7ex\hbox{E}\kern-.125emX}}
\begin{document}

\title{Bias-Aware Loss for Training Image and Speech Quality Prediction Models from Multiple Datasets}

\author{
\IEEEauthorblockN{
Gabriel Mittag$^1$, Saman Zadtootaghaj$^1$, Thilo Michael$^1$, Babak Naderi$^1$, Sebastian M\"oller$^{1,2}$
}
\smallskip
\IEEEauthorblockA{
$^1$Quality and Usability Lab, Technische Universit\"at Berlin, Berlin, Germany \\
$^2$Deutsches Forschungszentrum f\"ur K\"unstliche Intelligenz (DFKI), Berlin, Germany \\ 
 {first.last@tu-berlin.de}
}
}

\maketitle

\begin{abstract}
The ground truth used for training image, video, or speech quality prediction models is based on the Mean Opinion Scores (MOS) obtained from subjective experiments. Usually, it is necessary to conduct multiple experiments, mostly with different test participants, to obtain enough data to train quality models based on machine learning. Each of these experiments is subject to an experiment-specific bias, where the rating of the same file may be substantially different in two experiments (e.g. depending on the overall quality distribution). These different ratings for the same distortion levels confuse
neural networks during training and lead to lower performance. To overcome this problem, we propose a bias-aware loss function that estimates each dataset's biases during training with a linear function and considers it while optimising the network weights. We prove the efficiency of the proposed method by training and validating quality prediction models on synthetic and subjective image and speech quality datasets.
\end{abstract}

\section{Introduction}
In order to optimise the Quality of Experience (QoE) of multimedia services, developers rely on measures to validate new codecs or communication channels. Traditionally, the quality of image, video, or speech services is measured in subjective experiments, in which test participants are asked to rate the quality of a sample. The average across all test participants' ratings gives the so-called Mean Opinion Score (MOS). However, because this procedure is time-consuming and costly, instrumental models have been developed to automatically estimate the quality. In the case of speech quality, full-reference models, such as PESQ and POLQA have been established. In the image quality domain, many different visual quality metrics have been proposed \cite{lin2011perceptual}. The most popular metrics are NIQE \cite{niqe} and BRISQUE \cite{BRISQUE} for no-reference assessment or PSNR and SSIM \cite{iqa_3} for full-reference assessment. More recently also deep learning approaches have been introduced for speech quality or synthesised speech naturalness \cite{mittag2021nisqa, Catellier2020, Dong20, Lo2019, mittag2020b} and for image quality \cite{bosse2017deep, Chen_attiq, ren2018ran4iqa} prediction.

While objective quality models always compute the same score for a given sample, the MOS determined through subjective quality experiments, which is used as ground truth for training such objective models, is a sensitive measure. Minor change in a vote given from one test participant leads to a change of the overall value \cite{naderi2020transformation}. Consequently, test-retest studies even with the same group of participants, who rate the same dataset, often do not lead to the exact same MOS values~\cite{naderi2020transformation,p1401}. Given that, it is recommended to consider each subjective experiment as a closed set \cite{p1401}.

Datasets with subjective MOS are usually limited in size due to the maximum time in which one participant is able to rate the corpus before fatigue occurs. Therefore, it is common practice to use multiple datasets for training deep-learning-based quality prediction models. Furthermore, subjective data is usually sparse due to the costs that experiments involve and thus also older datasets are included for model training to increase the training size. Consequently, as these datasets often come from different labs with different test participants, and often, many years lie in between them, they are exposed to dataset-specific, subjective biases that make a comparison of the MOS values from different experiments difficult. The main bias-inducing factors according to ITU-T Rec. P.1401 \cite{p1401} are:\footnote{Besides these factors, \cite{zielinski2008on} lists further biases that occur in  listening tests.}
\begin{itemize}

    \item \textbf{Rating noise} The score assigned by a listener is not always the same, even if an experiment is repeated with the same samples and the same presentation order.
    
    \item \textbf{Order-effect} Subjects are influenced by the short-term history of the samples they previously rated. For example, after one or two poor samples, participants tend to rate a mediocre sample higher. In contrast, if a mediocre sample follows high-quality samples, there is a tendency to score the mediocre sample lower. Because of this effect, the presentation order for each subject is usually randomised in quality experiments.
    
    \item \textbf{Corpus-effect} The largest influence is given by effects associated with the average quality, the distribution of quality, and the occurrence of individual distortions. Test participants tend to use the entire set of scores offered in an experiment. Because of this, in an experiment that contains mainly low-quality samples, the subjects will overall rate them higher, introducing a constant bias. Despite verbal category labels, subjects adapt the scale to the qualities presented in each experiment. Furthermore, individual distortions that are presented less often are rated lower, as compared to experiments in which samples are presented more often, and people become more familiar with them. For example, it was shown in \cite{moller2006impairment} that a clean narrowband speech signal obtains a higher MOS in a corpus with only narrowband conditions than in a mixed-band corpus that includes wideband conditions as well.
    
    \item \textbf{Long-term dependencies} These effects reflect the general cultural behaviour of the subjects as to the exact interpretation of the category labels, the cultural attitude to quality, and language dependencies. Also, the daily experiences with telecommunication or media are important. Quality experience, and therefore expectation, may change over time as the subjects become more familiar with high-definition codecs and VoIP distortions.
    
\end{itemize}

In this paper, \textit{bias} refers to offsets and gradients between datasets that are caused by these factors. While offsets are mostly introduced by the overall quality that is presented to the participants during an experiment, gradients are, for example, introduced when the experiment does not cover the entire quality range (i.e. the ratings tend to become more pessimistic faster). 

These induced biases result in different MOS values for stimuli of the same distortion levels if they are contained in a different dataset. As a consequence, when multiple datasets are combined for training a quality prediction model, the correct rank order of MOS values is no longer given. The error-prone rank order of the combined training set leads to lower prediction performance of the trained quality prediction model. To overcome this problem, usually a set of common anchor conditions are included in all datasets (typically 20\% of test conditions). These anchor conditions cover the whole range of quality distortions and make a comparison of different datasets possible. By calculating a mapping function between the anchor conditions of the different datasets, biases can be partly removed before training. However, if the datasets that are used for training originate from different sources, they often do not contain the same anchor conditions. It should further be noted that biases that are introduced by the corpus-effect cannot be removed by normalising the datasets, which is a common preprocessing step. In case two datasets contain a different range of distortion levels, normalisation of the MOS values may even increase the bias between datasets. 

So far in literature, all deep learning based image or speech quality prediction models (e.g. the aforementioned models in \cite{mittag2021nisqa, Catellier2020, Dong20, Lo2019, mittag2020b, bosse2017deep, Chen_attiq, ren2018ran4iqa}) do not consider these biases between datasets and apply either a vanilla MSE (mean squared error) or MAE (mean absolute error) loss function. 

In this paper, we present a method that deals with subjective biases between datasets in the training phase of deep neural network models without the need for anchor conditions. The novelty of the proposed loss function is that it learns the biases automatically by using the model predictions as objective, unbiased measure. The proposed bias-aware loss is repeatedly updated during the model training, which avoids an overfitting to the biases themselves. The presented algorithm in this paper together with the synthetic bias experiments are made publicly available.
\footnote{https://github.com/gabrielmittag/Bias-Aware-Loss} 

The rest of the paper is structured as follows: At first we present the proposed loss function and the corresponding algorithm. The method is then evaluated on a synthetic dataset for which the biases are known. After that the bias-aware loss is applied to speech and image quality datasets with subjective MOS ratings.
\section{Method}
The basic idea is that the dataset-specific biases are learned and accounted for 
during the training of the neural network. After each epoch, the predicted MOS values of the training data are used to estimate the bias in each dataset by mapping their values to the ground truth subjective MOS values. The hypothesis is that the predicted MOS values of the model are objective in the sense that they will average out the biases of the different datasets while training. 

The biases are approximated with a first-order polynomial function. After the biases are estimated for each dataset, they can be used in the next epoch to calculate the proposed bias-aware loss. To this end, the predicted MOS values are mapped with the calculated bias coefficients. The MSE between the bias-mapped predicted MOS and the subjective MOS values then gives the loss as follows:\footnote{In this paper, we use MSE as the base loss function, however, the bias-aware loss algorithm can be applied to any distance measurement function.}

\begin{equation}
\label{eq:loss}
    l = \mathcal{L}(\mathbf{{y}, \widehat{y}, b}) 
     = \frac{1}{N} \sum^N_{i=1}  \left( y_i - ( b_0^j + b_1^j \widehat{y}_i ) \right)^2 
\end{equation}
where $i$ is the index of the sample belonging to dataset $j$, $y_i$ is the subjective MOS and $\widehat{y}_i$ is the predicted value for that sample, $b_0^j$ and $b_1^j$ are the estimated bias coefficients for dataset $j$, and $N$ is the overall number of training samples.

Because the predicted values are mapped according to the bias of each dataset, errors between the predicted and subjective values that only occur due to the dataset-specific bias are neglected. The model thus learns to predict quality, rather than non-relevant biases. 
\subsection{Learning with bias-aware loss}
The complete algorithm of the bias-aware loss is depicted in Algorithm \ref{alg:algorithm}. The inputs to the algorithm are the input features $x_i$ and the subjective MOS values that are the desired output values $y_i$ for all samples $i \in N$. Further, a list that contains the dataset index $j$ of each sample $\mathrm{db_i}$ is needed to assign the individual samples to their datasets. 
Before the training starts, the bias coefficients $b^j$ of all datasets are initialised with an identity function, with which Eq. \eqref{eq:loss} corresponds to a vanilla MSE loss. Because the model will typically not give a meaningful prediction output after the first few epochs, the $\mathrm{update\_bias}$ flag is set to False until a predefined model accuracy $r_{\mathrm{th}}$ in terms of Pearson's correlation coefficient (PCC) is achieved (see analysis in Sect~\ref{sec:results_synth}). Until this threshold is not reached the bias coefficients will not be updated, and therefore, a vanilla MSE loss is used for calculating the loss. As a metric PCC is used instead of RMSE (root-mean-square error) as the RMSE is strongly affected by biases and therefore unsuitable in this case.

\begin{algorithm}[ht!]
\caption{Training with bias-aware loss function}
\label{alg:algorithm}
\textbf{Input}: $x_i$: input features, $y_i$: subjective MOS values, $\mathrm{db_i}$: list of dataset indices $j$ for each sample $i$\\
\textbf{Parameter}: $r_\mathrm{th}$: minimum prediction accuracy to update the bias coefficients \\
\textbf{Output}: model weights
\begin{algorithmic}[1] 
\STATE{Number of dataset: $D = \mathrm{max(\mathbf{db})}$}
\STATE{Initialise bias for each dataset: $b^j=[0, 1], j \in D$}
\STATE{$\mathrm{update\_bias=False}$}
\WHILE{not converged}
\STATE{Shuffle mini-batch indices $\mathrm{idx}_k$}
\STATE $k=0$
\FOR{\ all mini-batches}
\STATE{Get mini-batch: \\ $\mathbf{x_b} = \mathbf{x}[\mathrm{idx}_k], \mathbf{y_b} = \mathbf{y}[\mathrm{idx}_k]$ } \label{alg:mini}
\STATE{Feed forward: $\mathbf{\widehat{y}_b} = \mathrm{model}(\mathbf{x_b})$}
\STATE Calculate bias-aware loss with Eq. \eqref{eq:loss}:\\ $l=\mathcal{L} (\mathbf{{y_b}, \widehat{y_b}}, b^j)$. \label{alg:loss}
\STATE Backpropagate \& optimise weights
\STATE $k=k+1$
\ENDFOR
\STATE{Predict MOS: $\mathbf{\widehat{y}} = \mathrm{model}(\mathbf{x})$}
\STATE{Calculate Pearson's correlation $r=\mathrm{PCC(\mathbf{y}, \mathbf{\widehat{y}}})$}.
\IF{$r >$ $r_\mathrm{th}$ \textbf{or} update\_bias}  \label{alg:rth}
\STATE $\mathrm{update\_bias=True}$
\FOR{$j$ in $D$}
\STATE{Find dataset indices:\\ $\mathrm{idx} = \mathrm{find(\mathbf{db} == \mathrm{j})}$} \label{alg:up1}
\STATE{Get dataset: $\mathbf{y^\mathrm{db}} = \mathbf{y}[\mathrm{idx}]$, $ \mathbf{\widehat{y}^\mathrm{db}} = \mathbf{\widehat{y}}[\mathrm{idx}]$, \\$\mathbf{M} = \mathrm{len(\mathrm{idx})}$ } \label{alg:db}
\STATE{Estimate bias: \\ $ \min_{b^j} \frac{1}{M} \sum^M_{i=1}  \left( {y}^{\mathrm{db}}_{i} - ( b_{0}^j + b_{1}^j {\widehat{y}}^{\mathrm{db}}_{i} ) \right)$ } \label{alg:up2}
\ENDFOR
\ENDIF
\ENDWHILE
\STATE \textbf{return} model weights
\end{algorithmic}
\end{algorithm}

At the start of each epoch, the mini-batch indices $\mathrm{idx}_k$ are randomly shuffled. It is necessary to preserve these indices in order to assign the samples to their corresponding datasets. After each epoch, the model is used to predict the MOS values $\mathbf{\widehat{y}}$ of all samples. Once the model accuracy $r_{\mathrm{th}}$ is reached, the $\mathrm{update\_bias}$ flag is set to $\mathrm{True}$ and the biases will be estimated after every epoch. 

To update the bias coefficients, the algorithm loops through all datasets individually, where $\mathrm{idx}$ represents the indices of all samples that belong to the dataset $j$. At line \ref{alg:db} of the algorithm, the subjective and predicted MOS values of the samples belonging to dataset $j$ are loaded. Then they are used to estimate the bias coefficients $b^j$. In the next training epoch, the biases are then applied to calculate the loss (line \ref{alg:loss}) of each mini-batch that may contain a random number of different datasets, and therefore each sample may also be subject to a different bias. By using the bias-aware loss, these biases are considered when calculating the error between the predicted and subjective MOS values. After each epoch, the bias coefficients are then updated to be in line with the updated model predictions. For more details on the algorithm see also the open-sourced PyTorch code.

\subsection{Anchoring Predictions}
\label{sec:bias_anchor}
When the bias-aware loss is applied, the MOS predictions are not anchored to the absolute subjective MOS values and, therefore, can wander off. While the predictions will still rank the samples in the best possible way, there may be a large offset between predictions and subjective values on the validation set. This effect will lead to a higher RMSE while the PCC is usually not affected. To overcome this problem, the predicted MOS values of all samples can be mapped to the subjective MOS after the training. This mapping can then be applied when making new predictions on validation data.

Alternatively, instead of estimating the bias for all datasets, the predictions can be anchored to one specific training dataset. This approach can be particularly useful if there is one dataset of which it is known that the conditions are similar to the conditions that the model should be applied to later. A new dataset is usually created with new conditions and then split into a training and validation set. To increase the training data size and to improve the model accuracy, older datasets may also be included in the training set. It is likely that there will be a bias between the new dataset and the older dataset, for example, because the highest quality in the new dataset will be higher than the one in the older datasets. The predictions can be anchored to the new dataset by omitting the bias update for the new dataset only (skipping line \ref{alg:up1}-\ref{alg:up2} in Algorithm \ref{alg:algorithm}). The biases of all other datasets are then computed in relation to the anchor dataset and, as a result, the final model predictions will be in line with the anchor dataset.

\section{Experiments and results}
In the first experiment, we generate a synthetic speech quality dataset for which the biases are known. After that, we conduct two more experiments with real speech quality and image quality data. Because the results of each neural network training run depend on random initialisation, random shuffling and other factors, such as random dropout, we run each experiment 15 times and use the average results to rule out any random effects.

\subsection{Synthetic data}
\label{sec:results_synth}
\begin{figure}[!ht]
\begin{minipage}[b]{.48\linewidth}
\centering
\centerline{\includegraphics[width=\linewidth]{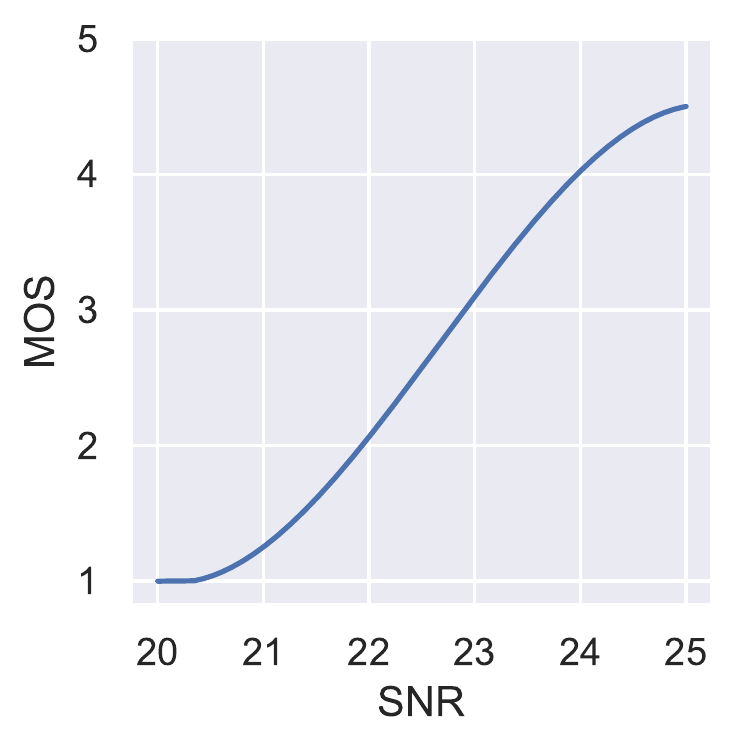}}
\caption{Mapping between noise in SNR and speech quality MOS.}
\label{fig:sigma2mos}
\end{minipage}
\hfill
\begin{minipage}[b]{0.48\linewidth}
\centering
\centerline{\includegraphics[width=\linewidth]{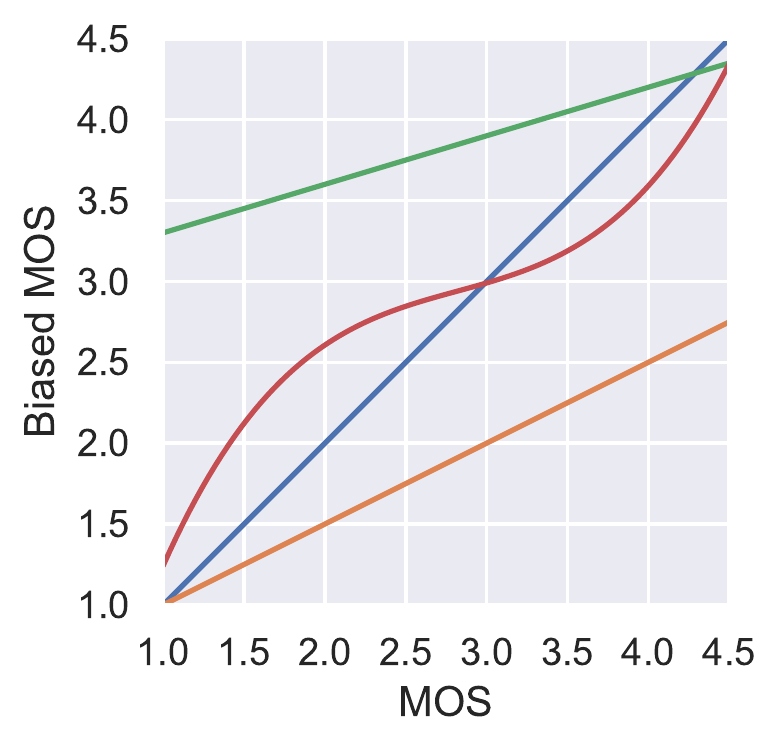}}
\caption{Four artificial biases introduced to the four simulated training datasets.}
\label{fig:biases}
\end{minipage}
\end{figure}

Firstly, a synthetic speech quality dataset is generated to which artificial biases are applied. As source files, the 2--3\,s reference speech files from the TSP dataset \cite{tspdb} are used. For the four training datasets overall 320 speech files and for the validation set 80 speech files are used. The speech signals were processed with white Gaussian noise to create conditions of different distortion levels. It was found that when the SNR range of the added noise is too wide, it is too easy for the model to predict the quality, and when it is too narrow, the prediction becomes too challenging. Therefore, the speech files were processed with noise at SNR values between 20\,dB and 25\,dB, which showed to be a good compromise. To simulate MOS predictions, an S-shaped mapping between the technical impairment factor (i.e., SNR) and MOS, taken from ITU-T Rec. G.107 \cite{ITUTRec.G.107} is used. The relationship is shown in Figure \ref{fig:sigma2mos} and maps the SNR values to a MOS between 1 and 4.5. 

The training data is divided into four different subsets, and a different bias is applied to each of them. These four artificial biases are shown in Figure \ref{fig:biases}. To better analyse the influence of the bias-aware loss, extreme biases are used in this experiment. There is no bias applied to the first simulated database (blue line). The second and the third simulated databases have linear biases applied, while the fourth database is exposed to a bias modelled with a third-order polynomial function. Each of the training datasets contains 80 files; the validation set also contains 80 files. The synthetic experiments are run by using the speech quality model NISQA \cite{mittag_a}.

\begin{figure}[!ht]
\begin{minipage}[b]{.48\linewidth}
\centering
\centerline{\includegraphics[width=\linewidth]{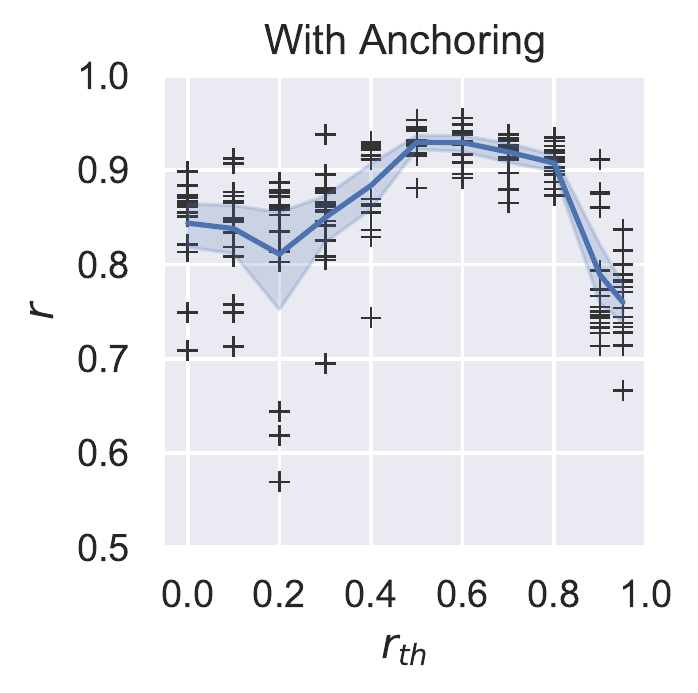}}
\caption{Validation results in terms of Pearson's correlation over 15 training runs with anchoring for different $r_{\mathrm{th}}$ thresholds.}
\label{fig:r_th_with_anchor}
\end{minipage}
\hfill
\begin{minipage}[b]{0.48\linewidth}
\centering
\centerline{\includegraphics[width=\linewidth]{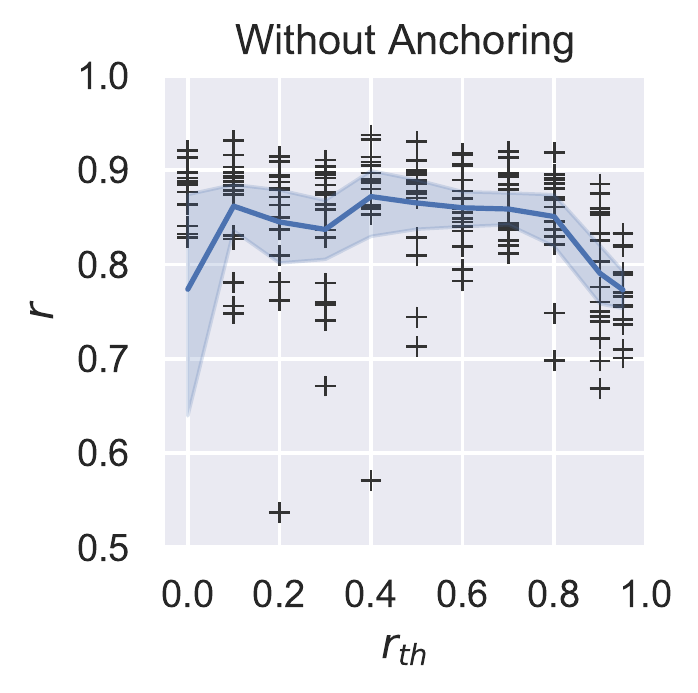}}
\caption{Validation results in terms of Pearson's correlation over 15 training runs without anchoring for different $r_{\mathrm{th}}$ thresholds.}
\label{fig:r_th_without_anchor}
\end{minipage}
\end{figure}

\textbf{Minimum Accuracy $r_{\mathrm{th}}$:} In the first experiment, the influence of the minimum PCC that must be achieved on the training set before activating the bias update in the bias-aware loss algorithm (line \ref{alg:rth} in Algorithm \ref{alg:algorithm}) is analysed. To this end, the experiment is run with 11 different threshold $r_{\mathrm{th}}$ between 0 to 0.95. An early stop on the validation PCC of 20 epochs is used, and the best epoch of each run is saved as result. The training run of each of these 11 configurations is repeated 15 times. The results, together with the mean results and their 95\% confidence interval, can be seen in Figure \ref{fig:r_th_without_anchor} for training without anchoring and in Figure \ref{fig:r_th_with_anchor} with anchoring. In the case of anchoring, the bias-aware loss is not used on the first dataset \textit{train\_1} but only on the other three datasets. 

The correlation of the results is highly varying when an anchor dataset is used. The highest correlation can be achieved for thresholds between 0.5 and 0.7. When no anchoring is applied, the exact threshold does not seem to be as crucial, as long as it is somewhere between 0.1 and 0.8. However, the PCC remains overall lower than the higher PCCs that can be achieved with an anchor dataset. For a threshold higher than 0.9, the accuracy notably drops. It can be assumed that at this point, the model weights are already optimised too far towards the vanilla MSE loss and cannot always profit from the late activated bias-aware loss.

\begin{figure*}[!ht]
    \centering
    \includegraphics[width=0.85\linewidth]{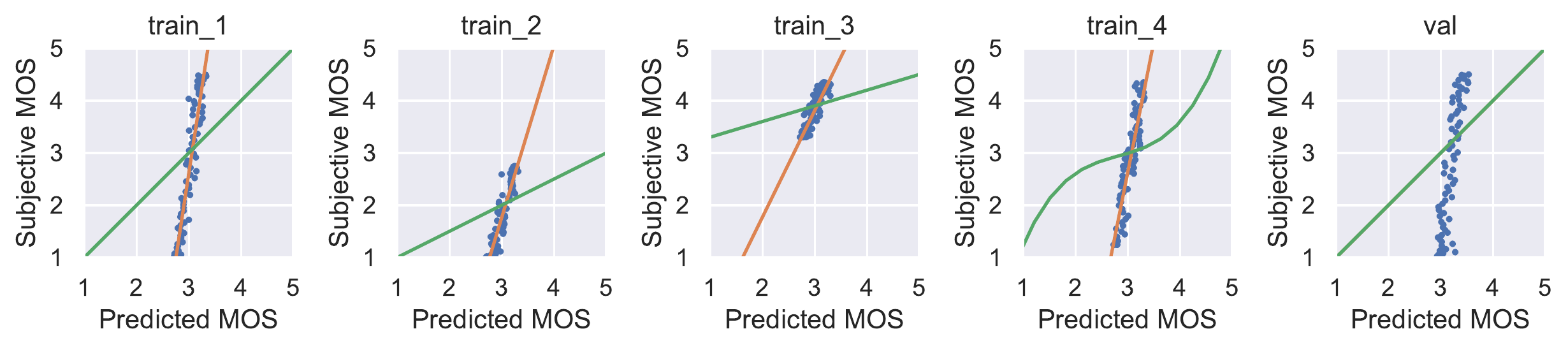}
    \includegraphics[width=0.85\linewidth]{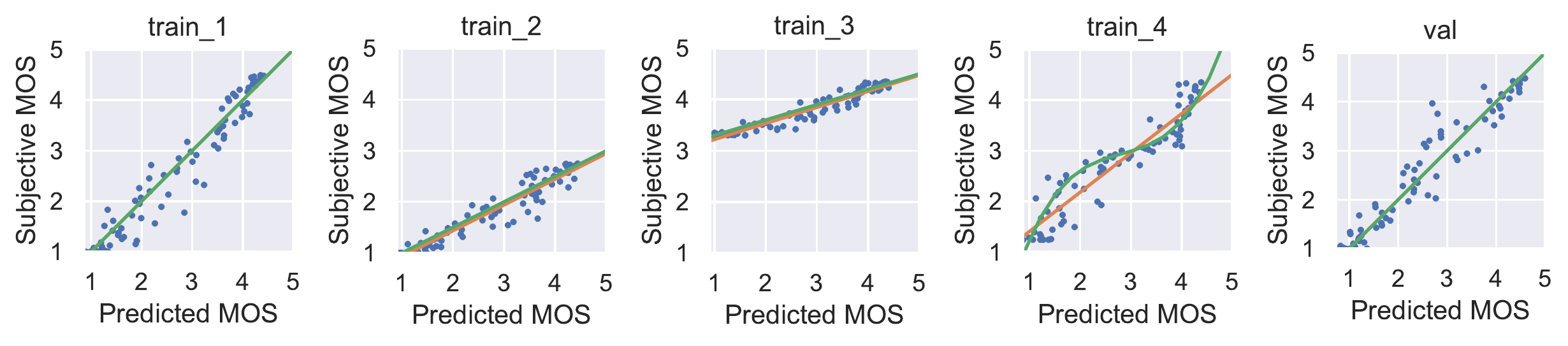}

    \caption{Two example training runs with bias-aware loss on the validation dataset. GREEN LINE: Introduced bias. ORANGE LINE: Estimated bias. 1st Row: Without anchoring. 2nd Row: Anchored with dataset train\_1.}
    \label{fig:synt_example}
\end{figure*}

\textbf{Synthetic training example:} Figure \ref{fig:synt_example} shows an example of four different training runs and different anchoring configuration. The figures show the epoch with the best results on the validation data set in terms of PCC. Each row presents one training run, and each column presents the results on the four different training datasets and on the validation dataset. The artificial bias that was applied to the datasets can be seen as green line (see also Figure \ref{fig:biases}). The estimated bias used by the bias-aware loss is depicted as orange line. The top row presents the results without anchoring and shows how the prediction results can drift away from the original values. While the predictions are extremely biased in this case, the achieved PCC remains high. 

The second row shows the results for anchoring with anchoring dataset. During the training, the bias of the first training dataset \textit{train\_1} was not estimated but fixed to an identity function. It can be seen that the prediction results on the validation set are less biased in this case. Furthermore, it can be seen that the model successfully learns the different biases when the estimated orange line is compared to the original green line.\footnote{We also experimented with third-order polynomial estimation functions with which biases such as in dataset \textit{train\_4} can be estimated more precisely. However, because the they did not further improve the results they are left out of this work.}

\begin{figure}[ht]
    \centering
    \includegraphics[width=0.6\linewidth]{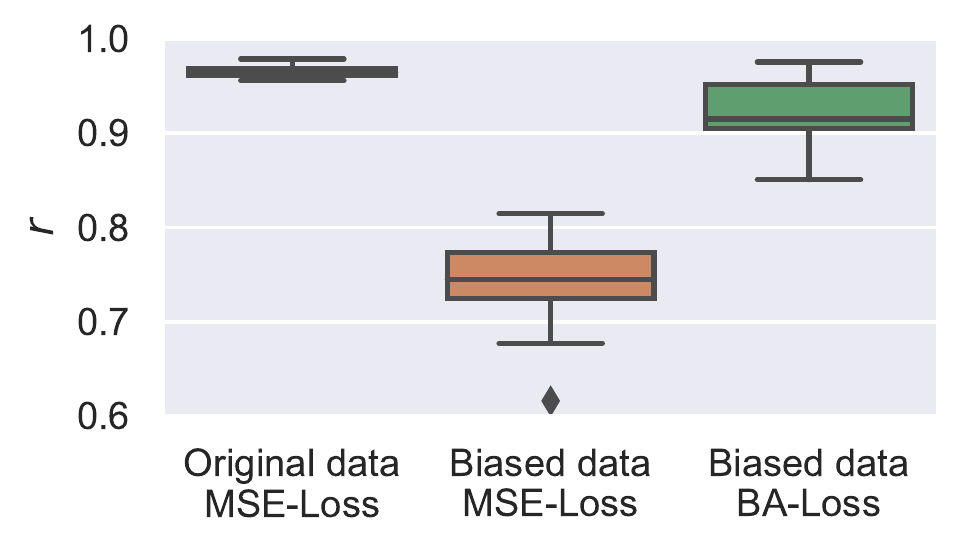}
    \caption{Validation results of 15 training runs on the synthesised data.}
\label{fig:iq_synt_config}
\end{figure}

\textbf{Results on synthetic data} The results on the validation dataset of 15 different runs are presented as boxplots in Figure \ref{fig:iq_synt_config}. When the second boxplot is compared to the first one where the original, unbiased data was used for training, it can be seen that the model accuracy decreases significantly from an average PCC of $r=0.95$ to $r=0.77$ when the training datasets are exposed to biases. This performance decrease caused by the biased training data can successfully be compensated for
by the bias-aware loss, which is displayed in the third boxplot on the right-hand side. The average PCC obtained is $r=0.93$, which is close to the results trained on unbiased data. Overall, the experiments show the efficiency of the proposed bias-aware loss when applied to the synthetic datasets, where the PCC can be increase from 0.77 to 0.93. 

\subsection{Speech Quality}

To analyse the efficiency on real datasets, the publicly available ITU-T Rec. P.Sup23 \cite{supp23} speech quality datasets with subjective quality ratings are used. The datasets originate from different sources and are therefore likely to be exposed to subjective biases. Six of the datasets that were rated on an ACR scale are used: EXP1a, EXP1d, EXP1o, EXP3a, EXP3c, and EXP3d. The proposed algorithm is analysed with a leave-one-dataset-out cross-validation, where the model is trained on the five remaining datasets. For each training run, the model of the epoch on which the best results in terms of PCC was achieved on the held-out validation dataset is saved. As prediction model again, the speech quality model NISQA is applied.

The training was run 15 times with an early stop of 20 epochs on the validation PCC, a learning rate of 0.001, and a mini-batch size of 60. The bias estimation was anchored to a randomly chosen dataset. The hyper-parameter $r_{\mathrm{th}}$ was optimised for each individual dataset. The average results for each dataset over all runs are shown in Table \ref{tab:results}. It can be noted that the efficiency of the proposed algorithm depends on the datasets it has been trained and evaluated on. For five of the six datasets, the proposed loss outperforms the vanilla MSE loss function, whereas, for dataset EXP1o, the vanilla MSE loss performs better. On the EXP3a dataset, a performance increase of about 0.05 in terms of PCC can be observed, showing that the speech quality prediction can notably be improved by applying the proposed bias-aware loss. 

\subsection{Image quality}
To analyse the efficiency on real image quality datasets, we use the following five publicly available datasets with subjective quality ratings, which are often used in image quality prediction research. It should be noted that the datasets come from different sources and each individual dataset may contain unique distortions that are not present in the other datasets. 

\textbf{CSIQ} \cite{csiq}: 866 images in total from 30 reference images, each distorted using one of five types of distortions. \textbf{TID 2013} \cite{tid}: 3,750 images in total from 24 reference images and 25 types of distortions. \textbf{Live Challenge} \cite{live_challenge_1}: 1,162 images with authentic image distortions captured using a representative variety of modern mobile devices. \textbf{Live IQA R2} \cite{iqa_1}: 779 images in total with 5 different distortion types. \textbf{Live MD} \cite{livemd_1}: 450 images in total, containing two types of multiply distorted images.

We again analysed the proposed algorithm with a leave-one-dataset-out cross-validation, where we trained the model on the four remaining datasets. As the datasets were rated on different scales, we linearly re-scaled all ratings to a range from 1--5, where 5 is the highest possible quality. Then, for each training run, we saved the model of the epoch on which the best results in terms of PCC was achieved on the validation dataset. As prediction model, we used Pytorch's ResNet50 implementation with 50 layers, where we replaced the last fully connected layer with a fully connected layer with only one output. Due to the large variety of different distortion in the datasets, we used the ImageNet pretrained weights and then fine-tuned the entire model during training. Because ResNet expects images of size 224x224x3, we only considered the center crop of the validation images to simplify the evaluation (instead of, e.g., averaging over multiple crops). However, to improve the accuracy of the predictions, a random crop on the training data was applied to increase the training data variety.

We ran the training 15 times with an early stop of 20 epochs on the validation PCC, a learning rate of 0.0001, and a mini-batch size of 32. The average results in terms of PCC are presented in Table \ref{tab:results}. On four of the five datasets, the proposed loss outperforms the vanilla MSE loss function, whereas, for dataset TID 2013, the vanilla MSE achieves the same results. The correlation of the Live Challenge dataset is overall very low with a PCC of 0.49/0.50. The low performance can be explained by the different types of distortions in this dataset (e.g., lens flare) compared to the training datasets. On the CSIQ dataset, a performance increase of 0.02 in terms of PCC can be observed, showing that the image quality prediction can be improved by applying the proposed bias-aware loss. 
\begin{table}[htb]
\centering
\captionof{table}{Validation results as average PCC of 15 training runs.}
\label{tab:results}
\begin{tabular}{@{}lcc@{}}
\toprule
Validation Dataset & MSE-Loss    & Proposed       \\ \midrule \midrule
Synthetic             & 0.77                & \textbf{0.93}  \\ \midrule
Speech Quality: EXP1a    & 0.95          & \textbf{0.96}      \\
Speech Quality: EXP1d    & 0.95          & \textbf{0.97}   \\
Speech Quality: EXP1o    & \textbf{0.96} & 0.95             \\
Speech Quality: EXP3a    & 0.87          & \textbf{0.92}            \\
Speech Quality: EXP3c    & 0.89          & \textbf{0.90}            \\
Speech Quality: EXP3d    & 0.89          & \textbf{0.92}               \\
\midrule
Image Quality: CSIQ                 & 0.82                & \textbf{0.85}  \\
Image Quality: Live Challenge       & 0.50         & \textbf{0.52} \\
Image Quality: Live IQA             & 0.88                   & \textbf{0.89}       \\
Image Quality: Live MD              & 0.81                 & \textbf{0.83}       \\
Image Quality: TID 2013             & 0.70          & 0.70   \\ 
\bottomrule
\end{tabular}
\end{table}

\section{Conclusion}
We presented an open-sourced loss function that automatically learns biases that occur when subjective quality experiments are conducted. The bias-aware loss does not punish prediction errors caused by biases, while the rank order within the dataset is predicted correctly.
We could show, on the basis of a synthesised dataset, that the proposed method obtains almost the same results as if the datasets were not exposed to any biases and outperformed the vanilla MSE loss by a relative improvement of 21\%. The performance on real data largely depends on the datasets that it is applied to. In cases where no biases are present between datasets, the model gives similar results as a vanilla MSE loss, however, in most cases the results could be improved. Because the bias-aware loss can be applied without additional data or computational costs, it is a helpful tool to improve speech, image, or video quality prediction models that are trained from multiple datasets.

\bibliographystyle{IEEEtran}
\bibliography{IEEEabrv,sample}

\end{document}